\documentstyle[proceedings]{article} 

\input{psfig.sty}
\input epsf

\def\s{\scriptstyle}
\def\ss{\scriptscriptstyle}

\def\ds{\displaystyle}

\newcommand{\be}{\begin{eqnarray}}
\newcommand{\ee}{\end{eqnarray}}
\newcommand{\nn}{\nonumber}

\def\slashxi{{\xi}\!\!\!/}

\def\o{\over}
\def\C{{\s C}}

\def\pmicl{P(\C_i,t+1)}

\def\pmicf{P'(\C_i,t)}
\def\pmicjf{P'(\C_j,t)}
\def\pmicc{P_c(\C_i,t)}
\def\pmicjc{P_c(\C_j,t)}

\def\pmicSf{P'(\C_i^{\ss \S}(m),t)}
\def\pmicCf{P'(\C_i^{\ss \c}(m),t)}

\def\pxl{P(\xi,t+1)}

\def\pxf{P'(\xi,t)}

\def\pxc{P_c(\xi,t)}
\def\pnxc{P_c(\slashxi_i,t)}

\def\pxSf{P'(\xi^{\ss \S}(m),t)}
\def\pxCf{P'(\xi^{\ss \c}(m),t)}

\def\fav{{\bar f}(t)}

\def\muti{{\cal P}({\ss C_i})}

\def\mutji{{\cal P}({\ss C_j\ra C_i})}
\def\mutis{{\cal P}({\s \xi})}
\def\mutijs{{\cal P}({\s \slashxi_i\ra\xi})}
\def\crossij{{\cal C}_{\ss C_iC_j}^{(1)}(m)}
\def\crossjl{{\cal C}_{\ss C_jC_l}^{(2)}(m)}

\def\hamij{d^H(i,j)}

\def\hamijS{d^H_{\ss\S}(i,j)}
\def\hamijC{d^H_{\ss\c}(i,j)}
\def\hamilS{d^H_{\ss\S}(i,l)}
\def\hamilC{d^H_{\ss\c}(i,l)}
\def\ra{\rightarrow}

\def\ef{f_{\ss\rm eff}(\xi,t)}

\def\e{{\rm e}}

\def\MM{{\cal M}}
\def\S{{\cal S}}
\def\c{{\cal C}}

\title{SOME EXACT RESULTS FROM A COARSE GRAINED FORMULATION OF 
GENETIC DYNAMICS} 
 
\author{ {\bf C. R. Stephens }\\ 
NNCP, Instituto de Ciencias Nucleares,\\   
UNAM, Circuito Exterior, A.Postal 70-543 \\ 
M\'exico D.F. 04510 \\
e-mail: stephens@nuclecu.unam.mx\\}

\begin{document} 
 
\maketitle

\begin{abstract} 

We extend a recently developed exact schema based, or coarse grained,
formulation of genetic dynamics \cite{stewael,stewael1,stewael2} 
and its associated exact Schema theorem 
to an arbitrary selection scheme and a general crossover operator.
We show that the intuitive``building block'' interpretation of the 
former is preserved leading to hierarchical formal solutions of the
equations that upon iteration lead to new results for the limiting
distribution of a population in the case of 1-point crossover and ``weak''
selection, where we define quantitatively ``weak''. We also 
derive an exact, analytic form for the population distribution 
as a function of time for a flat landscape and 1-point crossover.  

\end{abstract}

\section{Introduction}

Developing a better qualitative and quantitative understanding of the theory 
of genetic dynamics, and thereby of Genetic Algorithms (GAs), remains
a challenging problem by any measure. The benefits of having a better 
understanding are almost too obvious to mention. However, it is worth 
emphasizing the scale of the task. Almost all known theoretical results are
for systems with a small number of loci and even in the case of very simple
landscapes very little is known as far as explicit rather than formal, 
or numerical results is concerned. In particular we are unaware of any
systematic schemes for developing approximate solutions such as perturbative
methods. 

Passing beyond canonical results, such as Holland's Schema theorem \cite{holland},
which is associated with an inequality for the dynamics rather than an equality,  
various exact evolution equations have been derived previously: 
Goldberg and Bridges \cite{goldbridge} wrote down exact equations for two-bit problems. 
Later these equations were extended to three and four-bit problems \cite{whitley}.
These equations allowed for an explicit analysis of string gains and losses.
Whitley and Crabbe \cite{whitcrab} also presented an algorithm for 
generating evolution equations for larger problems that was equivalent 
to the earlier equation of \cite{goldbridge}.
Although exact these equations are extremely unwieldy and it is difficult to
infer general conclusions from their analysis. 

Another related approach is that
of Vose and collaborators \cite{vose,vosee,vose3} 
that treats GA evolution as a Markov chain. Noteworthy, and less familiar in the
GA community, is analogous work in mathematical evolution theory
\cite{slatkin,karlin} (see also the significant article of Altenberg 
\cite{altenberg} that has a foot in both camps).  
This formulation of GA dynamics appears to be so removed from elements such as 
the Schema theorem and Building Block hypothesis that the latter have very much
fell out of favour with aficionados of this dynamical formulation. Also, being
fundamental, microscopic equations they do not lend themselves easily to a 
treatment of schemata. It is important to note that this is not just an academic
point: genetic dynamics with a large number of degrees of freedom is so complex
that a formalism that treats the effective degrees of freedom at a more 
macroscopic level is certainly required. Such a formalism could be postulated 
directly, such as in the effective theory of Shapiro and collaborators \cite{shapiro},
or better, derived from the underlying microscopic dynamics. 

Evolution equations that offer the benefit of a very intuitive interpretation,
that illuminate the content of the Schema theorem and the Building Block hypothesis,
that naturally coarse grain from string equations to schema equations,
that yield an interpolation between the microscopic and the macroscopic 
and that offer new exact results or simpler proofs of known results have been
recently derived \cite{stewael,stewael1,stewael2}.  
These equations lead to many insights into the dynamics of GAs offering an
exact Schema theorem that naturally incorporates a form of the Building Block
hypothesis. Originally developed for a canonical GA (proportional selection, 1-point
crossover and mutation) the basic elements have also recently been extended to 
Genetic Programming (GP) by Poli and coworkers \cite{poli1,poli2}
who have showed the utility of the formalism by deriving several interesting 
new results.

In this paper we generalize the formalism of \cite{stewael,stewael1,stewael2}, 
to arbitrary selection schemes and a general crossover operator
showing that the basic advantages of the former are preserved. We see
that evolution can be fruitfully viewed as the hierarchical construction of
more and more complex building blocks. The usefulness of crossover 
in aiding search is governed by linkage disequilibrium coefficients associated
with selection probabilities rather than gene frequencies. 
We consider the limit distributions of a population showing that in the 
long time limit an arbitrary schema reaches Robbins proportions but with
a coefficient different than one. We also derive an exact, analytic form for 
the dynamics of a GA for a flat landscape and 1-point crossover.

\section{Microscopic String Evolution Equations}

In this section we recapitulate the results of \cite{stewael,stewael1,stewael2},
taking the opportunity to show their explicit generalization to 
a general crossover operator and arbitrary selection.
We start with an exact evolution equation that relates string 
proportions, $P(\C_i,t)=n(\C_i,t)/n$, for a genotype $\C_i$ consisting of 
binary alleles, at time $t$ to those at $t+1$, where
$n(\C_i,t)$ is the number of strings $\C_i$ in the population at time $t$
and $n$ is the population size. In the limit $n\ra\infty$ the $P(\C_i,t)$ 
give the probability distribution for the population dynamics and satisfy  
\be
\pmicl=\muti\pmicc\nn\\
+ \sum_{\ss \C_j\neq \C_i}\mutji\pmicjc\label{maseqtwo}
\ee 
where, for an infinite population, $\pmicc$ is the expected proportion 
of strings of type $\C_i$ after selection and crossover. 

The effective mutation coefficients are:
$\muti=\prod_{\s k=1}^{\ss N}(1-p_m(k))$, which is the probability that 
string $i$ remains unmutated, and $\mutji$, the probability that string $j$ 
is mutated into string $i$ given by 
\be
\mutji=\prod_{\ss{k\in \{C_j-C_i\}}}{p_m(k)}\prod_{\ss{k\in\{C_j-C_i\}_c}}
{(1-p_m(k))}
\ee
where $p_m(k)$ is the mutation probability of 
bit $k$. For simplicity we assume it to be constant, though the equations are 
essentially unchanged if we also include a dependence on time. 
$\s\{C_j-C_i\}$ is the set of bits that differ between $\C_j$ and 
$\C_i$ and $\s\{C_j-C_i\}_c$, the complement of this set, is the set 
of bits that are the same. In the limit where $p_m$ is uniform, 
$\muti=(1-p_m)^{\s N}$ and 
$\mutji=p_m^{\s d^{\ss H}(i,j)}(1-p_m)^{\s N-d^{\ss H}(i,j)}$,
where $\hamij$ is the Hamming distance between the strings $\C_i$ and $\C_j$.
Note that for a finite population 
the left hand side of (\ref{maseqtwo}) is the expected proportion of 
genotype $\C_i$ to be found at $t+1$ while any $P(\C_i,t)$ on the right 
hand side are to be considered as the actual proportions found at $t$.

Explicitly $P_c(\C_i,t)$ is given by
\be
\pmicc = \pmicf\nn\\
-{\sum_{\ss m=1 }^{\ss 2^N}}p_c(m){\ds\sum_{\ss C_j\neq C_i}}
\crossij\pmicf\pmicjf\nn \\ 
+{\sum_{\ss m=1 }^{\ss 2^N}}p_c(m)\sum_{\ss C_j\neq C_i}\sum_{\ss C_l\neq C_i}
\crossjl\pmicjf P'(\C_l, t)\label{eqnpc}
\ee
where ${\sum_{\ss m=1 }^{\ss 2^N}}$ is the sum over all possible 
crossover masks $m\in \MM$, $\MM$ being the space of masks, 
and $p_c(m)$ is the probability to implement the mask $m$. 
$\pmicf$ is the probability that genotype $\C_i$ is selected
and the coefficients $\crossij$ and $\crossjl$,  
represent the probabilities that, given that $\C_i$ 
was one of the parents, it is destroyed by the crossover process, and
the probability that given that neither parent was $\C_i$ it 
is created by the crossover process. 

In order to write these probabilities
more explicitly we denote the set of alleles inherited
by an offspring from parent $\C_j$ as $\S$ and the alleles inherited from  
parent $\C_l$, i.e. the set $\C_l-\S$, by $\c$. Naturally, $\S$ and $\c$
both depend on the particular crossover mask chosen. Then,  
\be
\crossij=\theta(\hamijS)\theta(\hamijC)\label{thetas}
\ee
and
\be
\crossjl={1\o2}[\delta(\hamijS)\delta(\hamilC)\nn\\
+\delta(\hamijC)\delta(\hamilS)]\label{deltas}
\ee
where $\hamijS$ is the Hamming distance between the strings $\C_i$ and
$\C_j$ measured only over the set $\S$, with 
the other arguments in (\ref{thetas}) and (\ref{deltas}) being similarly 
defined. $\theta(x)=1$ for $x>0$ and is $0$ for $x=0$, whilst 
$\delta(x)=0\ \ \forall x\neq0$ and $\delta(0)=1$. Note that $\crossij$ 
and $\crossjl$ are properties of the crossover process itself and therefore 
population independent. The equations 
(\ref{maseqtwo}) and (\ref{eqnpc}) yield an exact expression for the 
expectation values, $n(\C_i,t)$, and in the limit $n\ra\infty$ yield the 
correct probability distribution governing the GA evolution for arbitrary 
selection, mutation and crossover. It takes into account exactly the 
effects of destruction and construction of strings and, at least 
at the formal level, is either a generalization of or is equivalent to 
other exact formulations of GA dynamics \cite{goldbridge,whitley,vose,vosee,vose3}

As an explicit example, consider 1-point crossover. In this case there are
only $N-1$ pertinent crossover masks labelled by the crossover point $k$.
Then, $p_c(m)=p_c/N-1$ for $m=k$, $k\in[1,N-1]$ and $p_c(m)=0$ otherwise. 
Also, $\S=L$ and $\c=R$ (or vice versa) where $L$ and $R$ refer to the 
parts of the string to the left and right of the crossover point respectively.
For 2-point crossover there are ${}^{\s N-1}C_2$ non-zero masks labelled 
by two crossover points, $k_1$ and $k_2$. Then $p_c(m)=p_c/{}^{\ss N-1}C_2$
for $m\in \{k_1,k_2\}$ with $k_1$, $k_2\in[1,N-1]$ and $k_1>k_2$.
In this case $\S$ represents the parts of the string outside $k_1$ and 
$k_2$ and $\c$ the part between them (or vice versa). As a final example, 
for uniform crossover $p_c(m)=p_c\mu^{N_{\ss\S}}(1-\mu)^{N-N_{\ss\S}}/2^N$
where $\mu$ is the probability that a given allele is inherited from parent
$\C_j$, $\S$ being the set of alleles inherited from $\C_j$.  

Computationally, the above representation is very redundant. The matrix 
$\crossjl$ has dimensionality $(2^N-1)\times (2^N-1)$ but only 
$(2^{N_{\ss\S}}-1)\times (2^{N_{\ss\c}}-1)$ matrix elements are non-zero where
$N_{\ss\S}$ is the number of alleles of $\C_i$ inherited from one parent 
and $N_{\ss\c}$ the number from the other. As a consequence, and also 
for other reasons, iterating the dynamics is exceedingly difficult. 
The equations (\ref{maseqtwo}) and (\ref{eqnpc}) offer little in terms 
of intuitive understanding of the evolutionary dynamics and, arguably,
not a great deal in terms of formal results, though Vose and coworkers
have made important contributions in this area. Intuitive elements of 
GA theory such as the Building block hypothesis seem to play no role here
and are not apparent in any degree.  
Additionally, a more general formulation giving the dynamics of schemata
is relatively unnatural in this formulation.  

\section{Coarse Grained Evolution Equations}

Far greater progress can be made by changing basis from a microscopic
string basis to a coarse-grained basis associated with schemata. Such
a basis emerges very naturally anyway from equation (\ref{eqnpc}).

To see this, consider first the destruction term. For a given crossover
mask the matrix (\ref{thetas}) restricts the sum over the strings $\C_j$ to 
those that differ from $\C_i$ in at least one bit both
in $\S$ and in $\c$. One can convert the sum over
$\C_j$ into an unrestricted sum by subtracting off those $\C_j$ that have 
$d^H_{\ss\S}(i,j)=0$ and/or $d^H_{\ss\c}(i,j)=0$. We then use the fact that
$\sum_{\s \C_j}\pmicjf=1$ to gain a tremendous simplification
of the destruction term. Similarly one may write unrestricted sums for the 
construction term by subtracting off explicitly those terms that have
been added. The unrestricted construction term then becomes
\be
\sum_{\ss C_j\supset C_i^{\ss\S}}\sum_{\ss C_l\supset C_i^{\ss\c}}
\pmicjf P'(\C_l, t)
\ee
where $\C_i^{\ss\S}$ is the part of $\C_i$ inherited from $\C_j$, i.e. the
alleles $\S$, while $\C_i^{\ss\c}$ is the part inherited from $\C_l$, 
i.e. the alleles $\c$. Obviously, both $\C_i^{\ss\S}$ and $\C_i^{\ss\c}$
are schemata. Noting that the extra terms originating from the conversion of the 
restricted sums to unrestricted sums in the destruction and construction
terms exactly cancel one can finally rewrite $P_c(\C_i,t)$ as
\begin{eqnarray}
P_c(\C_i,t)=\pmicf(1-p_c)\nn\\
+{\sum_{\ss m=1}^{\ss 2^N}}p_c(m)
\pmicSf\pmicCf
\label{stringfin}
\end{eqnarray}
where $p_c={\sum_{\ss m=1 }^{\ss 2^N}}p_c(m)$ is the probability
to implement crossover (irrespective of the mask) and 
\be\pmicSf=\sum_{\ss C_j\supset C_i^{\ss\S}}\pmicjf\ee
and similarly for $\pmicCf$. In the absence of mutation 
$P_c(\C_i,t)=P(\C_i,t+1)$. It is important to note here that in this
form the evolution equation shows that crossover explicitly introduces
the idea of a schema and the consequent notion of a coarse graining 
relating a string $\C_i$ to its building blocks,
$\C_i^{\ss\S}$ and $\C_i^{\ss\c}$, which are schemata of order $N_{\ss\S}$ and 
$N_{\ss\c}=N-N_{\ss\S}$ respectively. Notice by going to this 
coarse-grained basis that the potentially $(2^N-1)\times 2^N$ 
destruction terms, $(2^N-1)$ being the number of terms in the restricted 
string sum and $2^N$ being the number of possible crossover masks, 
have been reduced to only one term and that for a given crossover 
mask the $(2^N-1)^2$ construction terms have been converted into one single
crossover term $\pmicSf\pmicCf$. Actually, this somewhat overstates
the case as the schema averages contain $(2^N_{\ss\S}-1)$ and $(2^N_{\ss\c}-1)$ 
terms respectively. However, at the very least the change in basis has 
explicitly eliminated the zero elements of $\crossjl$.

So let's recapitulate how the three different operators - mutation, crossover
and selection - enter in this exact dynamics. First of all, selection
is embodied in the factor $P'$ which is a ``black box'' for the purposes
of the dynamics. That is to say, the highly simplified and elegant form of 
the dynamical equation (\ref{stringfin}) is independent of the type of 
selection used and is a direct consequence of the form invariance of $P'$
under a coarse graining. In other words the relationship between $P'(\C_i,t)$ at the
string level and $P'(\C_i^{\ss\S},t)$ at the coarse grained level is strictly
linear, i.e. $P'(\C_i^{\ss\S},t)=\sum_{\C_i\supset \C_i^{\ss\S}}P'(\C_i,t)$.
However, if one wishes to know how the functional dependence
on $P$ dynamically changes then one must implement a particular selection
scheme. For example, with proportional selection 
$P'(\C_i,t)=(f(\C_i)/{\overline f}(t))P(\C_i,t)$ where $f(\C_i)$ is the fitness
of string $\C_i$ and ${\overline f}(t)$ is the average population fitness. It
is easy to demonstrate in this case that 
$P'(\C_i^{\ss\S},t)=(f(\C^{\ss\S}_i,t)/{\overline f}(t))P(\C^{\ss\S}_i,t)$,
where $f(\C^{\ss\S}_i,t)$ is the fitness of the schema $\C^{\ss\S}_i$,
which demonstrates the form invariance of proportional selection. 
Thus, the form invariance in this case also holds at the level of the $P$s not
just the $P'$s. This will not generally be the case. For example with a 
``non-linear'' selection scheme where $P'(\C_i,t)=a(\C_i,t)P(\C_i,t)
+b(\C_i,t)P^2(\C_i,t)$ there is no natural form invariance at the level of
the $P$s as $P'(\C^{\ss\S}_i,t)\neq a(\C^{\ss\S}_i,t)P(\C^{\ss\S}_i,t)+
b(\C^{\ss\S}_i,t)P^2(\C^{\ss\S}_i,t)$.
We also see that the elegance of the equation (\ref{stringfin}) is 
independent of the type of crossover implemented. The details of the crossover
mechanism enter in $p_c(m)$ and also determine which alleles are included in
$\S$. 

To iterate the equation it
is clear that we need to know the dynamics of the schemata $\C_i^{\ss\S}$
and $\C_i^{\ss\c}$. Thus we need to obtain an equivalent equation for an 
arbitrary schema, $\xi$ of order $N_2$ and defining length $l$. 
To do this we need to sum with 
$\sum_{\C_i\supset\xi}$ on both sides of the equation (\ref{maseqtwo}). 
This can simply be done to obtain \cite{stewael,stewael1,stewael2}
\be
\pxl=\mutis\pxc + \sum_{\s \slashxi_i}\mutijs\pnxc\label{maseqthree}
\ee
where the sum is over all schemata, $\slashxi_i$, that differ
by at least one bit from $\xi$ in one of the $N_2$ defining bits of $\xi$. 
In other words any schema competing with $\xi$ and belonging to the same 
partition. $\pxc=\prod_{k=1}^{N_2}(1-p_m(k))$ is the probability that $\xi$ 
remains unmutated and $\pnxc=\prod_{k\in\{\xi-\slashxi_i\}}p_m(k)
\prod_{k\in\{\xi-\slashxi_i\}_c}(1-p_m(k))$ is the probability that the 
schema $\slashxi_i$ mutate to the schema $\xi$. 
$P_c(\xi,t)=\sum_{\C_i\supset\xi}P_c(\C_i,t)$ is the probability of finding
the schema $\xi$ after selection and crossover. Note the form invariance
of the equation. i.e. (\ref{maseqthree}) has exactly the same form as 
(\ref{maseqtwo}). To complete the transformation to schema dynamics we
need the schema analog of (\ref{stringfin}). This also can be obtained by
acting with $\sum_{\C_i\supset\xi}$ on both sides of the equation. One obtains
\be
\pxc= (1-p_c{N_{\ss{\MM}_r}(l,N_2)\over N_{\ss\MM}})\pxf \nn\\
+ {\sum_{\ss m\in {\MM}_r(l,N_2)}}p_c(m)\pxSf\pxCf\label{schemafin}
\ee
where $\xi^{\ss \S}$ represents the part of the schema $\xi$ inherited 
from the first parent and $\xi^{\ss \c}$ that part inherited from the second. 
$N_{\ss{\MM}_r}(l,N_2)$ is the number of crossover masks that affect $\xi$, ${\MM}_r$
being the set of such masks. $N_{\ss\MM}$ is the total number
of masks with $p_c(m)\neq0$. Obviously these quantities depend on the type 
of crossover implemented and on properties of the schema such as defining
length. 
 
Once again for the purposes of illustration we may consider some specific
examples. For 1-point crossover: $N_{\ss\MM}=N-1$, $N_{\ss{\MM}_r}=l-1$, 
${\sum_{\ss m\in {\MM}_c }}p_c(m)\ra {p_c/N-1}\sum_{k=1}^{l-1}$, where $k$
is the crossover point, and $\xi^{\ss \S}=\xi^L(k)$ the part of $\xi$
to the left of the crossover point and $\xi^{\ss \c}=\xi^R(k)$ is the part to
the right. Substituting into (\ref{schemafin}) one finds the results of
\cite{stewael,stewael1,stewael2}. For $m$-point crossover: 
$N_{\ss\MM}={}^{\ss N-1}C_m$,
$N_{\ss{\MM}-r}=({}^{\ss N-1}C_m-{}^{\ss N-l}C_m)$ and  
${\sum_{\ss m\in {\MM}_r }}p_c(m)\ra {p_c}\sum_{\ss k_1>k_2>...>k_m}$,
where $k_1, ..., k_m$ are the $m$ crossover points.

\section{Schema Theorem}

With the exact evolution equations in hand we can extend the exact Schema
theorem of \cite{stewael,stewael1,stewael2} to a general crossover operator and arbitrary 
selection scheme. Once again we state it through the concept of effective
fitness \cite{stewael,stewael1,stewael2}
\par\noindent{\bf Exact ``coarse grained'' Schema theorem} 
\be
P(\xi,t+1)={\ef\o \fav}P(\xi,t)
\ee
where
\be 
\ef = \Big(\mutis\pxc  \nn\\
+\sum_{\s \slashxi_i}\mutijs\pnxc\Big)
{\fav\o P(\xi,t)}
\ee
and $\pxc$ is given by equation (\ref{schemafin}). The interpretation of this
equation is as for its analog for 1-point crossover and proportional selection
\cite{stewael,stewael1,stewael2}: that schemata of higher than average effective fitness
are allocated an ``exponentially'' increasing number of trials over time. 
An intuitive version of the Building Block hypothesis is part and parcel
of this Schema theorem: (\ref{schemafin}) explicitly shows the competition
between schema destruction and construction and shows how a schema is
constructed from its building blocks $\xi^{\ss\S}$ and $\xi^{\ss\c}$ which
in their turn are composed of building blocks $\xi^{\ss\S,\S}$ and
$\xi^{\ss\S\c}$ and $\xi^{\ss\c,\S}$ and $\xi^{\ss\c\c}$ respectively, which
in their turn... The
particular building blocks depend of course on the particular mask. 
Note that the crossover dependent part of (\ref{schemafin}) for a given
crossover mask can be written as 
\be 
\Delta(\xi,m)=\pxf-\pxSf\pxCf
\ee
which is a selection weighted linkage disequilibrium coefficient and, 
other than in a flat landscape, is a more relevant quantity than the 
standard linkage disequilibrium coefficient which is associated with
the $P$s rather than the $P'$s. If $\Delta>0$ then crossover has a negative
effect, i.e. schema destruction dominates while if $\Delta<0$ crossover
is positive. Thus, whether the selection probabilities for the building blocks
of a given schema are correlated or anticorrelated determines the nature
of crossover. It is important to note that the hierarchical structure 
inherent in the iterative solution of (\ref{stringfin}) relating a string 
or schema to its more coarse grained antecedents terminates after $N_2$
steps when one arrives at the maximally coarse grained effective degrees
of freedom - 1-schemata. These play a priviliged role as they
cannot be destroyed by crossover and in the continuous time limit satisfy 
$P(\xi^{(i)},t)={\rm exp}\int_0^t((f(\xi,t')/\fav)-1)dt'$.

One of the strengths of the present coarse grained formulation is that,
as we shall see, much can be deduced simply by inspection of the basic 
formulas. We shall first of all put the basic equation into a yet more 
elegant form. We introduce a $2^N$-dimensional population vector, ${\bf P}(t)$,
whose elements are $P(\C_i,t)$, $i=1,...,2^N$. Equation (\ref{maseqtwo}) 
can then be written in the form
\be
{\bf P}(t+1)={\overline{\bf W}}{\bf P}_c(t)    
\ee
where the mutation matrix ${\overline{\bf W}}$ is real, symmetric and 
time independent and has elements
${\overline{\bf W}}_{ij}=p_m^{d^H(i,j)}(1-p)^{N-d^H(i,j)}$, where for 
simplicity 
we restrict now to the case of uniform mutation (the generalization to 
non-uniform mutation is straightforward). Restricting attention to the
case of selection schemes linear in $P(\C_i,t)$, ${\bf P}_c(t)$ can be written
as
\be
{\bf P}_c(t)={\overline{\bf F}}(t){\bf P}(t)+
\sum_{m=1}^{2^N}p_c(m){\bf j}(m,t)\label{veceqn}
\ee
where the selection - crossover destruction matrix, ${\overline{\bf F}}(t)$, 
is diagonal,
and takes into account selection and the destructive component of crossover.
Explicitly, for proportional selection ${\overline{\bf F}}_{ii}(t)
=(f(\C_i)/\fav)(1-p_c)$. Finally, the ``source'' matrix is given by
${\bf j}(m,t)=\pmicSf\pmicCf$. Defining the 
selection-crossover destruction-mutation matrix 
${\overline{\bf W}}_s(t)={\overline{\bf W}}{\overline{\bf F}}(t)$ we have
\be
{\bf P}(t+1)={\overline{\bf W}}_s(t){\bf P}(t)
+\sum_{m=1}^{2^N}p_c(m){\overline{\bf W}}{\bf j}(m,t)\label{matrixeqn}
\ee
The interpretation of this equation is that ${\bf j}(m,t)$ is a source 
which creates strings (or schemata) by bringing building blocks together
while the first term on the right hand side tells us how the strings
themselves are propagated into the next generation, the destructive
effect of crossover renormalizing the fitness of the strings. 

\section{Formal Solutions of the Basic Equations}

Needless to say solutions of these dynamical equations are hard to come by.
They represent, for binary alleles, $2^N$ coupled non-linear difference 
equations, or in the continuous time limit - differential equations. As shown
in \cite{stewael,stewael1,stewael2}, compared to a representation based on (\ref{eqnpc}),
even a formal solution of (\ref{stringfin}) in the absence of mutation and
for 1-point crossover and proportional selection yields much valuable 
qualitative information, such as 
a simple proof of Geiringer's theorem \cite{geir} and an extension of it to the weak
selection regime. Here we extend this formal solution to the case of general
crossover and mutation and for any selection scheme linear in $P(\C_i,t)$. 
The iterated solution of (\ref{matrixeqn}) is
\be
{\bf P}(t)=\prod_{n=0}^{t-1}{\overline{\bf W}}_s(n){\bf P}(0)\nn\\
+\sum_{m=1}^{2^N}p_c(m)\sum_{n=0}^{t-1}\prod_{i=n+1}^{t-1}{\overline{\bf W}}_s(i)
{\overline{\bf W}}\ {\bf j}(m,n)\label{solnnodiag}
\ee
This solution actually holds true for arbitrary schemata. The only changes
are that the vectors are of dimension $2^{N_2}$, the matrices of dimension
$2^{N_2}\times2^{N_2}$, the sum over masks for the construction terms
is only over the set $\MM_r$ and that the building blocks in $j(m,t)$ are
those of the schema rather than the entire string.  

The interpretation of (\ref{solnnodiag}) follows naturally from
that of (\ref{matrixeqn}). Considering first the case without mutation, 
the first term on the right hand side gives us the probability for propagating
a string or schema from $t=0$ to $t$ without being destroyed by crossover.
In other words $\prod_{n=0}^{t-1}{\overline{\bf W}}_s(n)$ is the Greens function
or propagator for ${\bf P}$. In the second term, $\widehat{\bf j}(m,n)$, 
each element is associated with the creation of a string or schema at time $n$
via the juxtaposition of two building blocks associated with a mask $m$. 
The factor $\prod_{i=n}^{t-1}{\overline{\bf W}}_s(i)$ is the probability 
to propagate the resultant string or schema without crossover destruction 
from its creation at time $n$ to $t$. The sum over masks and $n$ is simply
the sum over all possible creation events in the dynamics. This formulation
lends itself very naturally to a diagrammatic representation that will be
discussed elsewhere. The role of mutation is to mix the strings or schemata
created in the aforementioned process.

So, what can be deduced from (\ref{solnnodiag})? First of all let's consider the 
case of a flat fitness landscape with no mutation; then 
$\widehat{\bf P}(t)={\bf P}(t)$ and ${\overline{\bf W}}$ is the unit matrix,
${\bf 1}$, and ${\overline{\bf W}}_s=(1-p_c){\bf 1}$. Then, for an arbitrary 
schema $\xi$
\be
{\bf P}(t)=(1-p_c{N_{\ss{\MM}_r}(l,N_2)\over N_{\ss\MM}})^t{\bf P}(0)\nn\\
+(1-p_c{N_{\ss{\MM}_r(l,N_2)}\over N_{\ss\MM}})^t
{\sum_{\ss m\in {\MM}_r(l,N_2) }}p_c(m)\nn\\
\times \sum_{n=0}^{t-1}
(1-p_c{N_{\ss{\MM}_r(l,N_2)}\over N_{\ss\MM}})^{-(n+1)}{\bf j}(m,n)
\ee
Now, obviously $\lim_{t\ra\infty}(1-p_c{N_{\ss{\MM}_r(l,N_2)}
\over N_{\ss\MM}})^t=0$ hence ${\bf P}(t)\ra 0$ as $t\ra\infty$ unless the 
summation over time leads to a cancellation of this damping factor. Given 
that the building block constituents of ${\bf j}(m,n)$ are associated
with damping factors $(1-p_c{N_{\ss{\MM}_r}(l',N'_2)\over N_{\ss\MM}})^t$
and $(1-p_c{N_{\ss{\MM}_r}(l-l',N_2-N'_2)\over N_{\ss\MM}})^t$ this can only
occur if there is no damping of the consituent building blocks and this only happens if
they are 1-schemata as $N_{\ss{\MM}_r}(1,1)=0$, i.e. you can't cut a 1-schema.
Given that it's a flat landscape $P(\xi^{(i)},t)=P(\xi^{(i)},0)$ where 
$P(\xi^{(i)},t)$ is the probability of finding the 1-schema $\xi^{(i)}$
corresponding to the bit $i$, hence the only term of ${\bf j}(m,t)$ which
gives a non-zero contribution when integrated is 
$\prod_{i=1}^{N_2}P(\xi^{(i)},0)$. Thus the limit distribution is
\be
P^*(\xi)=\lim_{t\ra\infty}P(\xi,t)=\prod_{i=1}^{N_2}P(\xi^{(i)},0)
\ee
which is Geiringer's theorem for a general crossover operator. 

The only role the type of crossover is playing here is how fast the transient 
corrections to the limit distribution die out. The damping is controlled by
$N_{\ss{\MM}_r}(l,N_2)$, hence the bigger it is the faster the corresponding
transient dies out. For example, for entire strings (or schemata of defining 
length $N$) for $m$ and $m'$-point crossover, where $m>m'$
\be
{N^{(m)}_{\ss{\MM}_r}(l,N_2)\o N^{(m')}_{\ss{\MM}_r}(l,N_2)}
=\left({1-{\Gamma(N-l+1)\Gamma(N-m)\over \Gamma(N-l-m+1)\Gamma(N)}\over
{1-{\Gamma(N-l+1)\Gamma(N-m')\over \Gamma(N-l-m'+1)\Gamma(N)}}}\right)>1
\ee
where $\Gamma(N)=(N-1)!$. Hence, we can quantify exactly how quickly 
$m$-point crossover transients die off relative to $m'$-point crossover 
transients. 
 
We can take this further and generalize Geiringer's theorem to non-flat fitness
landscapes for selection schemes linear in $P(\C_i,t)$ such as proportional
selection and for 1-point crossover and without mutation. Without loss of
generality we write $P'(\xi,t)=(1+\epsilon\delta(\xi,t))$ where $\epsilon$
will serve as a control parameter to determine how ``weak'' selection is, 
$\epsilon\delta(\xi,t)$ being the selection pressure for the schema $\xi$.
Weak selection will imply $\epsilon|\delta(\xi,t)|<1\ \forall \xi,\ t$.
A 1-schema evolves as
\be 
P(\xi^{(i)},t)=\e^{\epsilon\Delta(\xi^{(i)},t,t')}P(\xi^{(i)},t')
\ee
where $\Delta(\xi^{(i)},t,t')=\int_{t'}^t\delta(\xi^{(i)},t'')dt''$ is the 
integrated selection pressure for the 1-schema.  

\par\noindent{\bf Theorem}

{\it The limit distribution, $P(\xi)^*$, of $P(\xi,t)$ as $t\ra\infty$ 
for ``weak'' selection and no mutation and 1-point crossover in the 
continuous time limit is}
\be
P(\xi)^*={\cal A}(\xi,N,l,p_c)\prod_{i=1}^{N_2}P(\xi^{(i)},0)\label{geirsel}
\ee
{\it where the selection dependent amplitude ${\cal A}(\xi,N,l,p_c)$ is given by}
\be
{\cal A}(\xi,N,l,p_c)=\nn\\
\lim_{t\ra\infty}
\int_0^{p_c(l-1)t\o N-1}\e^{-\tau}\e^{\left(1-{p_c(l-1)\o N-1}\right)
\Delta(\xi,t,t-{N-1\o p_c(l-1)})}\times\nn\\
\prod_{i=1}^{N_2}(1+\delta(\xi^{(i)},t-{\ss{N-1\o p_c(l-1)}}))
\e^{\sum_{i=1}^{N_2}\Delta(\xi^{(i)},t-{N-1\o p_c(l-1)},0)}d\tau
\ee
Note that for a flat landscape ${\cal A}\ra 1$. Due to space limitations 
here a proof of this theorem will be presented elsewhere, though we can outline 
the logic. The key to the theorem is deciding just how weak selection must 
be in order that all higher order schema transients die out. It is for this 
reason that a tuning parameter was introduced. For a given weak selection 
landscape and values of $l$, $N$ and $p_c$ there exists a critical value of 
$\epsilon$, $\epsilon_{cr}(\xi,N,l,p_c)$, below which all non-1-schema transients
die out as $t\ra\infty$. One examines the growth or decay rate of a schema
and all its constituent building blocks. Growth is caused by selection and
decay by crossover. By examining the dominant growth mode, other than that 
associated with pure 1-schemata, one can tune $\epsilon$ such that it becomes
a decay mode. All other growing modes are subdominant and therefore also 
decay. Hence, the only non-zero mode in the limit $t\ra\infty$ is the 
1-schema mode of (\ref{geirsel}). The difference with the result for a flat
landscape is that in this case the effective fitness landscape for 1-schemata 
is not flat. As an example, if one considers a weak counting ones landscape, i.e.
where $f(\C_i)=1+\epsilon n_1(\C_i)$, $n_1(\C_i)$ being the number of ones
on the string, then $\epsilon_{cr}\sim p_c/N^3$ for strings. 

\section{Explicit Solutions}

To show the further utility of the present formalism we will find the general
explicit solution to (\ref{stringfin}) for a flat fitness landscape, without mutation
and for 1-point crossover in the continuous time limit. Lest this be thought 
a trivial problem it's wise to remember that it still involves the solution
of $2^N$ coupled non-linear differential equations. To illustrate the general
principles we'll consider first a three bit problem. The general structure
of the equations (\ref{stringfin}) is that one builds up a solution via intermediate 
building blocks. The most fundamental blocks are 1-schemata as these cannot
be cut and hence transform trivially under recombination (save in the case
of uniform crossover). ${\bf P}^{(l,N_2)}(t)$ in this case is an 
$N$-dimensional vector and ${\bf P}^{(l,N_2)}(t)={\bf P}^{(l,N_2)}(0)$,
with $l=1$ and $N_2=1$. 
Explicitly, $P(i**,t)=P(i**,0)$, $P(*j*,t)=P(*j*,0)$ and $P(**k,t)=P(**k,0)$.
There are four 2-schemata per crossover point corresponding to 
$P(ij*,t)$ and $P(*jk,t)$, $i,\ j =0,\ 1$. $P(ij*,t)$ satisfies 
\be
P(ij*,t)=\e^{-{p_ct\o2}}P(ij*,0)\nn\\
+{p_c\o 2}\e^{-{p_ct\o2}}
\int_0^t P(i**,t') P(*j*,t')\e^{{p_ct'\o2}}dt'\label{twoschema}
\ee
with an analogous equation for $P(*jk,t)$. The simple solution of 
(\ref{twoschema}) is
\be
P(ij*,t)=\e^{-{p_ct\o2}}P(ij*,0)\nn\\
+(1-\e^{-{p_ct\o2}})P(i**,0)P(*j*,0)\label{twoschemasoln}
\label{twosol}
\ee
The 3-schema solution is found using (\ref{twoschemasoln})
\be
P(ijk,t)=\e^{-{p_ct}}P(ijk,0)\nn\\
+{p_c\o2}\e^{-{p_ct}}
\int_0^t \e^{{p_ct'}} \left(P(ij*,t')P(**k,t')\right.\nn\\
\left.+P(i**,t')P(*jk,t')\right)dt'
\label{threeschema}
\ee
Substituting the solution (\ref{twosol}) into (\ref{threeschema}) one finds
simply
\be
P(ijk,t)=\e^{-{p_ct}}P(ijk,0)\nn\\
+\e^{-{p_ct\o2}}(1-\e^{-{p_ct\o2}})\left(P(ij*,0)P(**k,0)\right.\nn\\+
\left.P(i**,0)P(*jk,0)\right)\nn\\
+(1-\e^{-{p_ct\o2}})^2 P(i**,0) P(*j*,0)P(**k,0)\label{threesol}
\ee
In the limit $t\ra\infty$, $P(ijk,t)\ra P(i**,0)P(*j*,0)P(**k,0)$. We see here
the approach to Robbin's proportions is exponentially fast with the bigger
schemata dying off quicker. 

The general solution for an $N$-bit string is
\be
P(\C_i,t)=\sum_{n=0}^{N-1}\e^{-{np_ct\o N-1}}(1-\e^{-{p_ct\o N-1}})^{N-n-1}
{\cal P}(n+1)\label{exactsol}
\ee
where ${\cal P}(n+1)$ is an initial condition and represents a partition over
the probabilities for finding $N-n$ building blocks at $t=0$. For a given 
$n$ there are ${}^{\ss N-1}C_n$ such terms. Equation (\ref{threesol}) offers a 
simple illustration where $N=3$, hence $n=0,\ 1,\ 2$. $n=0$ corresponds to the
$N$ building block terms of which there are ${}^{\ss N-1}C_0=1$ term, 
$P(i**,0)P(*j*,0)P(**k,0)$. $n=1$ corresponds to the
$N-1$ building block terms of which there are ${}^{\ss N-1}C_1=2$ terms, 
$P(ij*,0)P(**k,0)$ and $P(i**,0)P(*jk,0)$. Finally, $n=2$ corresponds to the
$N-2$ building block terms of which there are ${}^{\ss N-1}C_2=1$ term,
$P(ijk,0)$.

Notice that (\ref{exactsol}) gives not only the asymptotic behaviour but
also the complete transients. We are unaware of any similar result in the
literature. It is not difficult to prove that (\ref{exactsol}) is the general
solution by showing that it satisfies (\ref{stringfin}). This requires the solutions
of (\ref{stringfin}) for $\C^L(k)$ and $\C^R(k)$, the building blocks of $\C_i$ also. 
From the form invariance of the equations the solution of (\ref{stringfin}) is
\be
P(\C^L_i,t)=\nn\\
\sum_{n_L=0}^{l_L(k)-1}\e^{-{n_Lp_ct\o N-1}}
(1-\e^{-{p_ct\o N-1}})^{N-n_L-1}
{\cal P}(n_L+1)\label{exactsolL}
\ee
where $N-n_L$ is the number of building blocks of $\C_i^L(k)$
and $l_L(k)$ is the defining length of $\C_i^L(k)$. Using the analogous
equation for $\C_i^R(k)$ and the fact that $n=n_L+n_R$ one sees that both
sides of (\ref{stringfin}) have the same time dependence hence it suffices to prove
the equivalence at $t=0$. The equality at $t=0$ hinges on the identity 
that ${\cal P}(n+1)=\sum_{n_L=0}^{n-1}{\cal P}(n_L+1){\cal P}(n_R+1)$.

In the case of zero mutation the other exact limit of the evolution 
equations is that of pure selection. Interestingly, the solution is only 
exact in the case of discrete time. With these two exact limits in hand
there should be scope for obtaining perturbative solutions to the evolution
equations either in powers of $\epsilon$ for weak selection or in powers of
$p_c$ for strong selection.

\section{Conclusions}

In this article we have generalized the formalism of \cite{stewael,stewael1,stewael2}
to cover arbitrary selection schemes and a general crossover operator. 
We have tried to emphasize the advantages of this formulation and in particular
show how these advantages have concrete payoffs. We believe that our coarse
grained formulation is more intuitive in its content than others, giving
an exact schema theorem that contains in a very obvious manner a Building Block 
hypothesis. Its coarse grained hierarchical structure, both in terms of 
time and building block complexity, leads to a formulation wherein results
such as Geiringer's theorem can be seen in such a manifestly simple way that
it is essentially proof by inspection. Additionally, the way that selection,
schema destruction and schema creation enter in different ways leads,
using the same hierarchical structure, to a version of Geiringer's theorem for
a class of ``weak'' selection landscapes where we were able to quantify the 
meaning of ``weak''. 

The fundamental coarse grained equation show an important
form invariance, i.e. after passing from schemata of a given $N_2$ to more 
coarse grained schemata of order $N'_2<N_2$, the resulting equations have exactly
the same form. This is certainly not manifestly true of the fundamental string
equations (\ref{eqnpc}). This form invariance is an important property as it implies 
that if a solution can be found for one degree of coarse graining then an
analogous solution can simply be written down for the more coarse grained degrees 
of freedom. We showed that the eficacy of the present formulation lies not
only in its intutive appeal and the facility with which more general formal
results can be deduced but also in how one may derive explicit analytic
results, as in the case of an exact solution for the evolution of strings  
in the case of a flat fitness landscape and 1-point crossover. 

We believe that our previously derived results and the results herein are the
tip of the iceberg and that the present formalism may serve as a starting
point for deriving a whole host of similar results and beyond. To name just a 
couple of obvious ones: finding the asympotic limiting distributions for other types
of crossover and including in mutation and finding the exact dynamics for other
types of crossover operator
One of the most intriguing possibilities, that we have briefly alluded to, is
the possibility of deriving approximate results using a systematic approximation
scheme such as perturbation theory. The existence of exact solutions in the
flat fitness landscape and the zero crossover limit add weight to such a 
supposition as does the fact that the iterative solution of the equations 
(\ref{stringfin}) leads to a diagrammatic series very similar to the Feynman
diagrammatic series that appear in quantum field theory.  

\subsubsection*{Acknowledgements} 

The author thanks Ken De Jong, Bill Spears and Lashon Booker for useful 
discussions and Christopher Ronnewinkel for comments on the manuscript.
 
\subsubsection*{}

\end{document}